\documentclass{emulateapj}
\usepackage{graphicx}
\usepackage{epsf}
\usepackage[z]{tmsfnts}
\usepackage{multirow}

\newcommand{\kev}{\rm{keV}}
\newcommand{\hmpc}{h^{-1}\, {\rm{Mpc}}}
\newcommand{\hkpc}{h^{-1}\, {\rm{kpc}}}
\newcommand{\Te}{T_{e}}
\newcommand{\Tion}{T_i}
\newcommand{\Tgas}{T_{\rm{gas}}}

\begin{document}
\bibliographystyle{apj}

\slugcomment{{\em The Astrophysical Journal Letters, accepted}} 
\shortauthors{RUDD \& NAGAI}
\shorttitle{TWO-TEMPERATURE ICM AND THE SZ}

%========================================================================
\title{Non-Equilibrium Electrons and the Sunyaev-Zel'dovich Effect of Galaxy Clusters}
%========================================================================

\author{Douglas H. Rudd}
\affil{School of Natural Sciences, Institute for Advanced Study, Princeton, NJ 08540 USA}
\and
\author{Daisuke Nagai}
\affil{Department of Physics, Yale University, New Haven, CT 06520}
\affil{Yale Center for Astronomy \& Astrophysics, Yale University, New Haven, CT 06520}

%---------------------------------------------------------------------
\begin{abstract}
We present high-resolution cosmological hydrodynamic simulations of three galaxy clusters 
employing a two-temperature model for the intracluster medium.  We show that electron 
temperatures in cluster outskirts are significantly lower than the mean gas temperature, because 
Coulomb collisions are insufficient to keep electrons and ions in thermal equilibrium.  
This deviation is larger in more massive and less relaxed systems, ranging from 5\% in 
relaxed clusters to 30\% for clusters undergoing major mergers.  The presence of non-equilibrium 
electrons leads to significant suppression of the SZE signal at large cluster-centric radius.  
The suppression of the electron pressure also leads to an underestimate of the hydrostatic 
mass.  Merger-driven, internal shocks may also generate significant populations of non-equilibrium 
electrons in the cluster core, leading to a 5\% bias on the integrated SZ mass proxy during cluster 
mergers.
\end{abstract}
%---------------------------------------------------------------------

\keywords{ galaxies:clusters:general - intergalactic medium }

%---------------------------------------------------------------------
\section{Introduction}
\label{section:intro}
%---------------------------------------------------------------------

In current theories of cosmological structure formation, the cold dark matter and baryon fluid 
undergoes gravitational collapse, leading to virialized filaments and roughly spherical "halos" 
on a variety of scales.  Collisionless shocks play a critical role in this process by converting the 
kinetic energy of in-falling baryonic material to thermal energy.  This heating leads to large 
reservoirs of $T=10^6-10^8\ \mathrm{K}$ gas that fills and surrounds massive structures,
directly tracing the formation and evolution of cosmic structure.

Observational probes of this gas are generally sensitive only to the electron component of the 
plasma.  Typically, theoretical studies assume that the electrons are in thermal equilibrium 
with the surrounding ions.  However, this is not a good approximation in the low-density 
outskirts of galaxy clusters due to the extended timescale for electrons to reach equilibrium via 
Coulomb collisions \citep{fox_loeb97,ettori_fabian98}.  In the pre-shock intergalactic 
medium, the bulk of the kinetic energy is carried by the heavier ions.  Since the electrons and 
ions are only weakly coupled, this leads to an electron temperature in the post-shock plasma 
that is significantly lower than that of ions.  The post-shock electrons and ions gradually 
relax to a single equilibrium temperature.  

The low emissivity of gas at densities of $10^{-4}-10^{-5}\ \mathrm{cm}^{-3}$ has heretofore 
limited X-ray spectroscopic temperature measurements to regions where electrons are expected 
to have reached equilibrium \citep{markevitch_etal98,vikhlinin_etal05,pratt_etal07}.  Advances
in the sensitivity of X-ray instruments are pushing the limit of ICM temperature measurements to 
a significant fraction of the virial radius \citep[e.g.,][]{george_etal09}.  Such measurements 
remain expensive with the current generation of instruments, however.

In recent years the thermal Sunyaev Zel'dovich effect (SZE) has been suggested as a possible 
probe of gas in this phase \citep[e.g.,][]{afshordi_etal07,hallman_etal07,hallman_etal09}, due 
to its weaker density dependence.  Observations using the next generation of SZ experiments 
are ongoing, and are expected to provide accurate measurements of the electron pressure beyond
the virial radius.  Large samples of massive clusters with profile measurements beyond the 
virial radius should therefore become available in the near future.

In this work we consider the impact of the two-temperature structure of the intracluster medium (ICM) 
on the thermal SZE.  We employ high resolution simulations of clusters which improve on previous studies 
\citep{takizawa99,chieze_etal98,courty_alimi04,yoshida_etal05,yoshikawa_sasaki06} by combining 
the efficient shock-capturing properties of Eulerian adaptive mesh refinement with high mass and spatial 
resolution such that both the large-scale accretion shocks and internal, merger-driven shocks are well 
captured.  We show that the lower electron temperature in the cluster outskirts leads to a significant 
underestimate of the gas pressure when derived through the SZE.  We also demonstrate that the two-temperature 
structure of the ICM is sensitive to recent accretion history through internal merger-driven shocks and 
discuss the relative importance of these shocks to the SZE.

%---------------------------------------------------------------------
\section{Two Temperature Cosmological Cluster Simulations}
\label{section:sims}
%---------------------------------------------------------------------

The simulations were performed using the ART code \citep{kravtsov_etal97,kravtsov_etal02}, 
which was modified to model a two temperature electron-ion plasma.  Following \citet{yoshida_etal05}, 
we divide the plasma into two components, electrons and ions, which are assumed to be individually 
in local thermodynamic equilibrium (LTE) with separate Maxwellian velocity distributions defined by 
temperatures $\Te$ and $\Tion$, respectively.  The separation into only two species, electrons 
and ions, each in separate equilibrium is reasonable, since the self-equilibration timescales for each 
specie, $t_{ii}$ and $t_{ee}$, are considerably shorter than the electron-ion equilibration timescale, 
$t_{ei}$ \citep{fox_loeb97,spitzer62}.\footnote{$t_{ei} \sim (m_i/m_e)^{1/2} t_{ii} \sim (m_i/m_e)t_{ee}$.}   
When an electron-ion plasma passes through a shock, most of the kinetic energy goes into heating 
the heavier ions, causing $\Tion >> \Te$.  After the shock, electrons and ions slowly equilibrate via 
Coulomb interactions, each converging to the mean gas temperature, 
$\Tgas = (n_e \Te + n_i \Tion)/(n_e+n_i)$, over a typical electron-ion equilibration 
timescale, $t_{ei}$.  The evolution of the electron temperature is given by,
\begin{equation}
\frac{dT_e}{dt} = \frac{\Tion - \Te}{t_{ei}} - (\gamma -1) T_e \left( \nabla \cdot \mathbf{v} \right),
\label{eq:dTdt}
\end{equation}
where the second term accounts for adiabatic compression heating and cooling. 
The timescale for equipartition between two charged species is given by \citet{spitzer62},
\begin{equation}
t_{eq} = \frac{3 m_1 m_2}{8 (2\pi)^{1/2} n_2 Z_1^2 Z_2^2 e^4 \ln \Lambda}\left(\frac{kT_1}{m_1} + \frac{kT_2}{m_2}\right)^{3/2},
\end{equation}
where $m$, $T$, and $Z$ are the mass, temperature, and charge of each specie, respectively, 
$n$ is the number density, and $\ln \Lambda \approx 40$ is the Coulomb logarithm.  For the 
fully ionized ICM, including contributions from both protons and He$^{++}$, the timescale for
equilibration is,
\begin{equation}
t_{ei} \approx 6.3 \times 10^{8} \mathrm{yr}\ \frac{\left(T_e/10^7 \mathrm{K}\right)^{3/2}}{\left(n_i/10^{-5} \mathrm{cm}^{-3}\right)\left(\ln \Lambda/40\right)}.
\label{eq:tei}
\end{equation}
Note that this timescale can be comparable to the Hubble time in regions with $T\sim 10^7$~K 
and overdensities $10-100$ with respect to the cosmic mean.  

While our model assumes negligible electron heating within shocks, some non-adiabatic
heating due to plasma instabilities is likely.  However, theoretical expectations for the amount 
and source of electron heating in these shocks vary widely \citep[see, e.g.,][for a recent
review]{bykov_etal08}, ranging from a constant fraction $\Te \sim 0.2\Tion$ of the pre-shock kinetic
energy \citep{cargill_papadopoulos88} to a constant post-shock electron temperature and
inverse-square scaling with shock velocity $\Te/\Tion \sim v_{s}^{-2}$ \citep{ghavamian_etal07}.
In either case we expect the electron heating in the high Mach number accretion shocks to 
be small.  There are some indications of more rapid electron equilibration in shocks with lower 
Mach number in the ICM \citep{markevitch06}, however, the process is not yet well constrained 
by observations to date.  We therefore neglect it in our model, noting that our results represent
a maximum of the possible effect.
  
%*********************************************************************
\begin{deluxetable}{lcccc}
\tablecolumns{5}
%\tablewidth{\linewidth} 
\tablecaption{List of Simulated Clusters\label{table:clusters}}
\tablehead{
\multirow{2}{*}{Name\tablenotemark{1}\hspace{0.05\linewidth}} &
\multicolumn{1}{c}{$r_{200}$} &
\multicolumn{1}{c}{$M_{200}$} &
\multicolumn{1}{c}{$T_m\tablenotemark{2}$} &
\multicolumn{1}{c}{Dynamical state\tablenotemark{3}} 
\\
& $(\hmpc)$ & $(10^{14} \hmpc)$ & $(\kev)$ & (Rel/Unrel)}
\startdata
CL101 \dotfill & 1.77 & 12.83                 & 6.40 & Unrelaxed\\
CL104 \dotfill & 1.42 & \phantom{0}6.69       & 5.44 & Relaxed \\
CL6   \dotfill & 0.95 & \phantom{0}1.98       & 2.14 & Relaxed
\enddata
\tablenotetext{1}{Cluster labels correspond to those used in \citet{nagai_etal07b}.}
\tablenotetext{2}{$T_m$ is the gas-mass weighted temperature measured within $r_{200}$.}
\tablenotetext{3}{Classification of dynamical state is described in \citet{nagai_etal07a}.}

\end{deluxetable}
%*********************************************************************

In this work we re-simulate three galaxy clusters selected from the sample presented in 
\citet[][N07 hereafter]{nagai_etal07b}.   The initial conditions and simulation parameters 
are identical to N07, except the simulations we present here neglect the physics of 
radiative cooling and star formation, which should have negligible impact in the cluster 
outskirts.  The cluster properties at $z = 0$ are summarized in Table~\ref{table:clusters}.  
These clusters are chosen to cover a range of mass and the dynamical histories.  CL101 is 
a massive, dynamically active cluster, which experiences violent mergers at $z\sim0.1$ and 
$z\sim0.25$.  CL104 is a similarly massive cluster, but with a more quiescent history. 
This cluster has not experienced a significant merger for the past 6~Gyrs, making it 
one of the most relaxed systems in the N07 sample.  The last cluster in our sample, CL6, 
is a fairly relaxed, low-mass cluster at $z = 0$, but experiences a nearly equal mass major 
merger at $z=0.6$.  Each cluster is simulated using a $128^3$ uniform grid with 8 levels of
refinement.  CL101 and CL104 are selected from $120\hmpc$ computational boxes and
CL6 is selected from a $80\hmpc$ box, achieving peak spatial resolution of $\approx3.6\hkpc$
and $\approx2.4\hkpc$, respectively.  The dark matter particle mass in the region surrounding
the cluster is $9\times 10^{8}\,h^{-1}\, {\rm M_{\odot}}$ for CL101 and CL104 and 
$3\times 10^{8}\,h^{-1}\,{\rm M_{\odot}}$ for CL6, while the rest of the simulation volume
is followed with lower mass and spatial resolution.

%---------------------------------------------------------------------
\section{Results}
\label{section:results}
%---------------------------------------------------------------------

%*********************************************************************
\begin{figure}[t]
%\plotone{f1}
\begin{center}\includegraphics*[width=0.95\columnwidth]{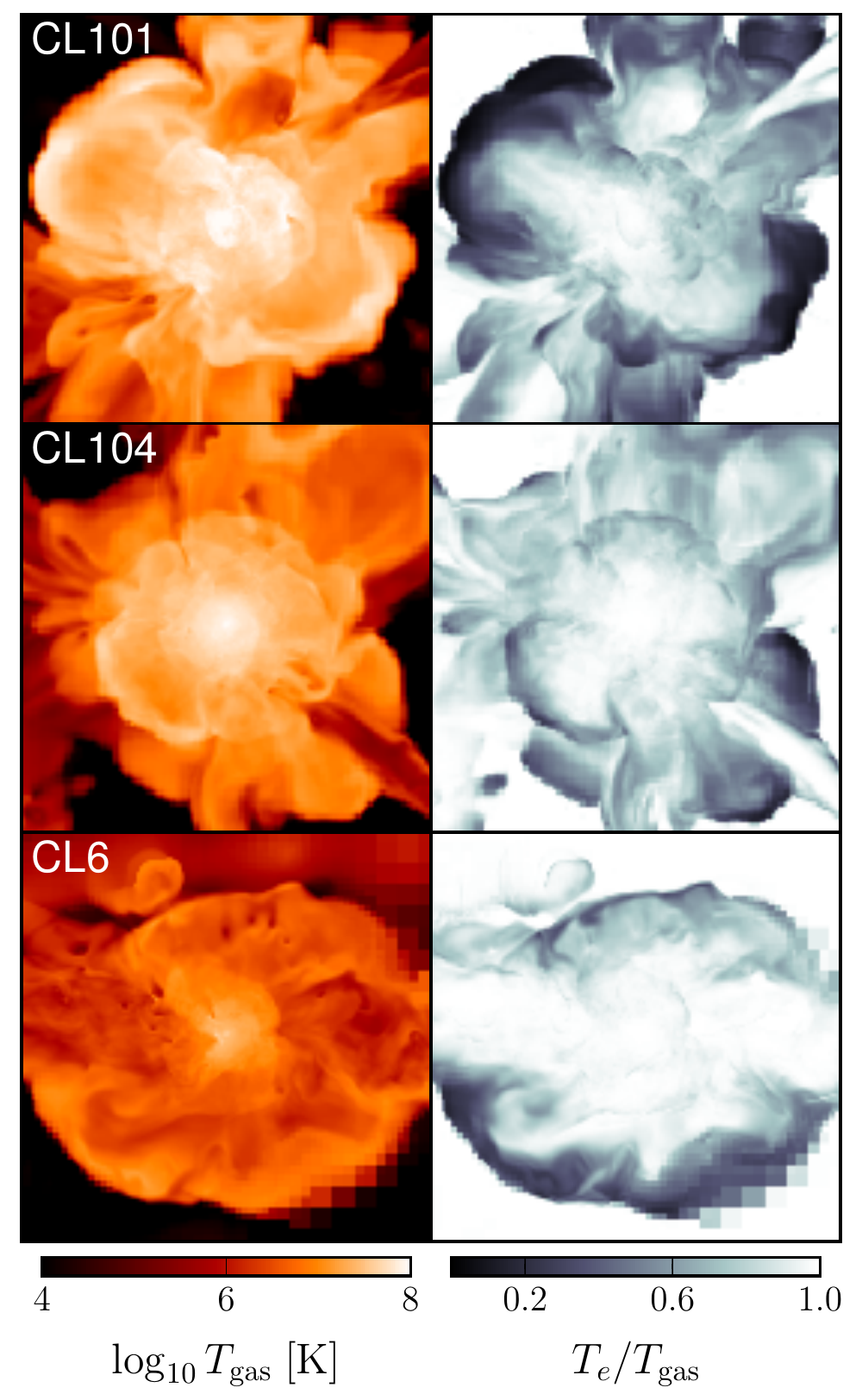}\end{center}
%\begin{center}\includegraphics*[height=0.8\textheight]{f1}\end{center}
\caption{The distribution of the mean gas temperature ($\Tgas$) and the ratio of electron and mean gas 
temperatures ($\Te/\Tgas$) for three simulated clusters at $z=0$.  {\it Left panel}: the projected mass-weighted 
gas temperature in $1\hmpc$ slices centered on each cluster and $12\hmpc$ on a side.  {\it Right panel}:
the ratio of the electron-temperature $\Te$ to the mean ICM temperature $\Tgas$.  
\label{figure:projected}}
\end{figure}
%*********************************************************************

Figure~\ref{figure:projected} illustrates the complex and highly aspherical distributions of 
gas temperature ({\it left panel}) and non-equilibrium electrons ({\it right panel}) in the 
outskirts of the simulated galaxy clusters at $z=0$.  These maps highlight a complicated network 
of non-equilibrium electrons associated with accreting and shocked material. The recent mergers 
in CL101 generate both internal and external shocks, and these shocks are responsible for the 
significant quantity of non-equilibrium electrons throughout the cluster.  Non-equilibrium 
electrons are found mostly in the cluster outskirts for the relaxed systems (CL104 and CL6), 
and electrons and ions in the inner regions are mostly in equilibrium.
 
%*********************************************************************
\begin{figure}[t]
\plotone{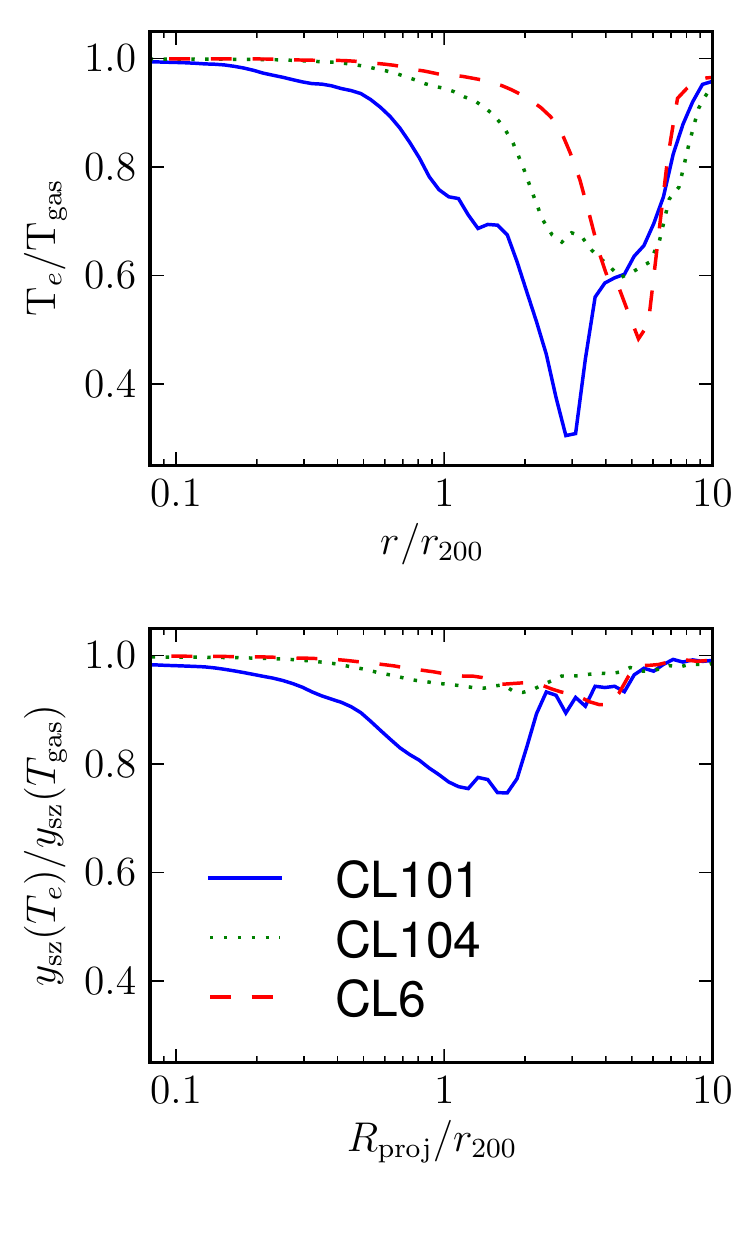}
%\begin{center}\includegraphics*[height=0.8\textheight]{f2}\end{center}
\caption{Profiles of ICM electron temperature relative to the mean ICM gas temperature 
for each cluster at at z = 0.  {\it Top panel}: 3D electron temperature profiles
plotted averaged in spherical shells and scaled to $r_{200}$.  
{\it Bottom Panel}: 2D mass-weighted electron temperature profiles averaged
in cylindrical annuli projected $60\hmpc$ through the simulation volume.  This quantity 
is equivalent to the bias in the observed SZ flux in each annulus with respect to the flux 
one would observe if the electron temperature, $\Te$, were equal to the mean gas temperature, 
$\Tgas$.  
\label{figure:profiles}}
\end{figure}
%*********************************************************************

The top panel of Figure~\ref{figure:profiles} shows the spherically averaged 3D radial profile 
of $\Te/\Tgas$ as a function of the cluster-centric radius for the simulated clusters at $z=0$.  
For the relaxed clusters, electrons and ions are in nearly perfect thermal equilibrium within 
$r=0.5 r_{200}$.  Beyond this radius, the electron temperature becomes increasingly smaller 
than the mean gas temperature, and the magnitude of the bias is inversely proportional to the 
ICM temperature: $\Te$ is biased low by 5\% at $r=r_{200}$ for CL104 
and $r=2 r_{200}$ for CL6.  The bias reaches its maximum at the virial shock, 
$r_{\rm shock}=[5-6] r_{200}$.  Beyond the shock radius, electrons and ions in the
unshocked gas are largely in equilibrium, with a slight departure from 
equilibrium due to the accretion shocks of large-scale filaments. 

%*********************************************************************
\begin{figure}[t]
\plotone{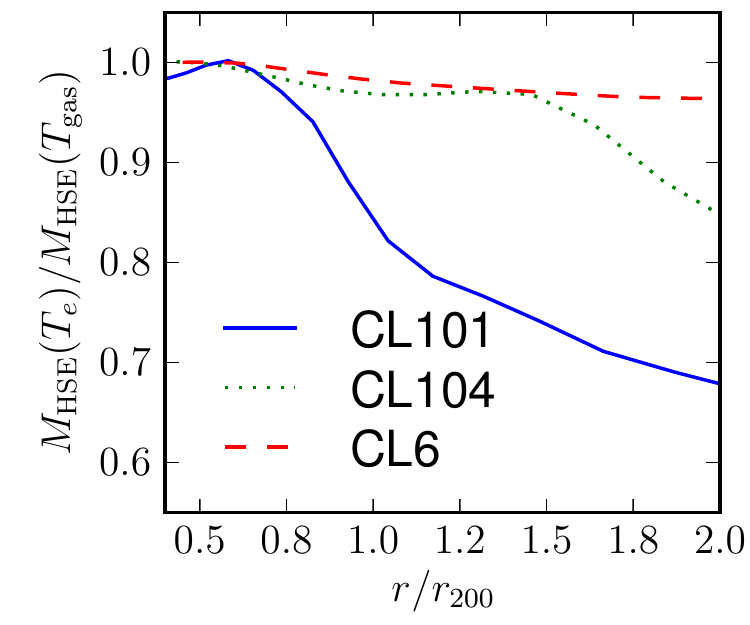}
\caption{The bias in the hydrostatic mass estimate of galaxy clusters due to the non-equipartition
of electrons and ions.  The bias is defined as the difference in the hydrostatic masses computed 
using the electron temperature and the mean gas temperature. 
\label{figure:hydrostatic}}
\end{figure}
%*********************************************************************

%*********************************************************************
\begin{figure*}
%\plotone{f4}
\begin{center}\includegraphics*[width=\textwidth]{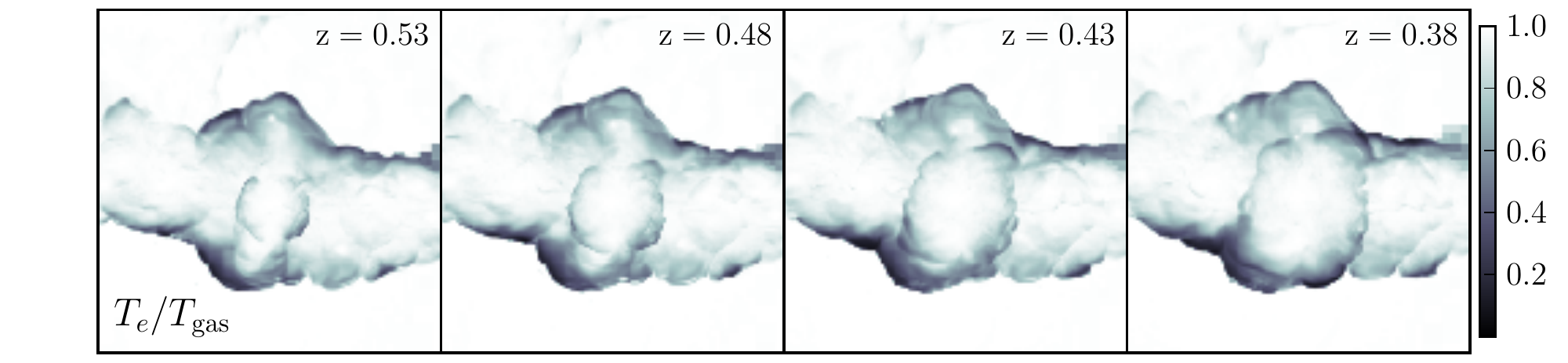}\end{center}
\caption{The evolution of $\Te/\Tgas$ surrounding $\mathrm{CL}\ 6$
from $z = 0.53$ to $z = 0.38$.  This cluster experiences a major merger 
at $z \sim 0.6$, which is nearly in the plane of the figure.  This 
merger causes symmetric merger shocks which propagate outward
from the center, eventually reaching and merging with the cluster virial
shock at $z \sim 0.3$.  
\label{figure:evolution}}
\end{figure*}
%*********************************************************************

The magnitude of the non-equipartition between electrons and ions depends sensitively
on the cluster dynamical state.  The largest deviation from thermal equilibrium is seen 
in the most massive and least relaxed system (CL101). For this cluster, we find 
$\Te/\Tgas \approx 0.75-0.9$ at relatively small radii $r=[0.5-1] r_{200}$ and reaching a 
minimum $\Te/\Tgas \approx 0.35$ at $r=3 r_{200}$.  The non-equilibrium electrons
at small radius are primarily generated by internal shocks associated with the recent
merger (see also Figure~\ref{figure:projected}).

Production of non-equilibrium electrons by internal shocks within the virialized regions of 
clusters is illustrated in Figure~\ref{figure:evolution}, which shows the evolution of $\Te/\Tgas$ 
for CL6 during the period immediately following a nearly equal mass merger around $z=0.6$ 
\citep[see also Figures~1 and 2 in][]{nagai_kravtsov03}.  The final coalescence 
generates a strong internal shock which propagates out from the cluster center exciting 
a thin, trailing layer of non-equilibrium electrons.  Although at fixed radius this 
is a transient phenomenon, lasting for only about $10^7-10^8\mathrm{yrs}$ due to the short equilibration 
timescale within the high-density ICM, the shock propagation occurs on a dynamical timescale.
This leads to non-equilibrium electrons in the cluster center for $0.5-1~\rm{Gyr}$ during the
merger.  These internal shocks can therefore introduce a significant temperature bias on considerably 
smaller scales than the accretion shocks.

The suppression of the electron temperature also leads to a bias in the hydrostatic mass estimate,
$M_{\rm HSE}(< r) \propto -dP/dr$, as shown in Figure~\ref{figure:hydrostatic}.  We follow the procedure
described in \citet{lau_etal09} for computing the hydrostatic mass profiles. Briefly, we reduce
spurious fluctuations in the numerical derivatives by removing gas in bound substructures and
then smooth the resulting profile with a Savitzky-Golay filter.  At small scales, $r \lesssim 0.7~r_{200}$, 
the deviation from the hydrostatic mass estimated  using $\Tgas$ is small for all three clusters, and 
remains $\lesssim5\%$ for the relaxed clusters out to $1.5~r_{200}$.  The bias in the hydrostatic 
mass is quite large for clusters undergoing recent major mergers (e.g., CL101), where the hydrostatic 
mass is underestimated by $20-30\%$ in the radial range $[1-2]~r_{200}$.  Note that this effect is 
in addition to the bias introduced by the non-thermal pressure due to random gas motions 
\citep{lau_etal09}.

These biases in temperature due to non-equilibrium electrons have important 
implications for the interpretation of Sunyaev-Zel'dovich effect (SZE) observations.  This is 
illustrated in the bottom panel of Figure~\ref{figure:profiles}, which shows that the presence of 
non-equilibrium electrons leads to significant suppression of the SZE signal at large cluster-centric 
radius.  We define the bias to be the ratio of the Compton-y parameters, defined as $y_{\rm SZ}(T) \equiv \int n_e T dl$, 
computed alternately using $\Te$ and $\Tgas$.  Again the bias is largest for the unrelaxed system, 
and is as large as $20\%$ in the radial range $[0.7-2]r_{200}$.  The bias is more moderate ($5-10\%$) 
for the other two relaxed systems in a similar radial range. 

The integrated Compton-y parameter, $Y_{\rm SZ}$ (or SZ flux), is also affected by the 
non-equilibrium electrons.  But, the biases are much smaller since the effect is averaged
over the entire cluster and is dominated by the central regions where electrons are in equilibrium.   
Examining the evolution of the bias in $Y_{\rm SZ}$ within the spherical region of $r_{500}$ for 
all three simulated clusters, we find the bias is no larger than 6\%, which occurs during periods
of major mergers (e.g., CL101 since the last major merger at $z\sim 0.1$ or CL6 at $z \sim 0.6$ 
as shown in Fig~\ref{figure:evolution}).  The bias reaches 9\% during the same period when measured 
within the larger radius, $r_{200}$.  This process may also affect the scatter in the mass-observable 
relations.  At various times during the last CL101 merger, however, the cluster lies both above and 
below the mean Y-M relation, while the bias due to non-equilibrium electrons always acts to suppress 
the SZ signal.  Large samples of simulated clusters are therefore needed to adequately address this 
effect. For comparison, we computed the bias in X-ray spectroscopic temperature using the 
{\it spectroscopic-like} formula of \citet{mazzotta_etal04}.  The resulting bias is negligible, 
even during mergers, since it is strongly weighted to gas at higher densities and correspondingly 
shorter $t_{ei}$.

%---------------------------------------------------------------------
\section{Conclusions}
\label{section:conclusions}
%---------------------------------------------------------------------

We use simulations of cosmological cluster formation to explore the two temperature structure 
of the ICM and its effects on the Sunyaev-Zel'dovich effect.  We show that electron temperatures 
are lower than the ion (or mean gas) temperature in the low-density outskirts of galaxy clusters, 
where Coulomb collisions are insufficient to keep electrons and ions in thermal equilibrium.  This 
leads to a decrease in the SZE relative to predictions which assume electron-ion equilibrium.  The 
suppression of electron pressure in turn leads to an underestimate of the hydrostatic mass.

Our simulations also show that the magnitude of the non-equipartition between electrons and ions 
depends sensitively on the cluster dynamical state.  The electron temperature (or pressure) 
is smaller than the mean gas temperature by about 5\% at $r=r_{200}$ for relaxed clusters, but 
the bias could be as large as 30\% for unrelaxed systems, in which populations of non-equilibrium 
electrons are created within the virial regions of cluster by internal, merger-driven shocks.  
Although these are short-lived in the high-density ICM, these non-equilibrium electrons can 
affect global properties, such as the integrated Compton-y or the mass weighted temperature, 
by about 5\%.   X-ray spectroscopic temperature, on the other hand, is not significantly affected 
even during the merger periods, since it is strongly weighted to gas at higher densities and 
correspondingly shorter $t_{ei}$.  These biases are the sources of systematic uncertainties that 
will need to be taken into account for the interpretation of upcoming large-scale SZE experiments, 
including ACT\footnote{Atacama Cosmology Telescope ({\tt http://www.physics.princeton.edu/act/})}, 
Planck\footnote{Planck Surveyor ({\tt http://planck.esa.int/})}, 
and SPT\footnote{South Pole Telescope ({\tt http://pole.uchicago.edu})}.

We lastly note that the details of electron equilibration will affect the ability
to detect virial shocks using the SZ \citep{kocsis_etal05,molnar_etal09} by smoothing
the electron temperature profile in the region of the shock.  Successful detections of
accretion shocks by ALMA\footnote{Atacama Large Millimeter Array ({\tt http://www.alma.nrao.edu/})} 
can be used to place strong constraints on the otherwise uncertain physics operating at these shocks.  
We will explore this possibility in an upcoming paper.

\vfill

\acknowledgements 
We thank Kevin Heng for helpful discussions during early stages of this 
work, Erwin Lau for providing his code to measure the hydrostatic mass, 
and Fang Peng for useful comments on the manuscript.
DHR gratefully acknowledges the support of the Institute for Advanced
Study.  The simulations were performed at the Joint Fermilab - KICP 
Supercomputing Cluster, supported by grants from Fermilab, Kavli Institute 
for Cosmological Physics, and the University of Chicago. 

%\bibliography{ms}

\end{document}